\documentclass[%
 reprint,
 amsmath,amssymb,
 aps,
]{revtex4-2}

\usepackage{graphicx}
\usepackage{dcolumn}
\usepackage{bm}
\usepackage{amsmath}

\begin{document}

\preprint{APS/123-QED}

\title{Correlated emission lasing in a single quantum dot embedded inside a bimodal photonic crystal cavity}

\author{Lavakumar Addepalli}
\email{d20034@students.iitmandi.ac.in}
\author{P.K. Pathak}
\email{ppathak@iitmandi.ac.in}
\affiliation{
 School of Physical Sciences, Indian Institute of Technology Mandi, Kamand, H.P., 175005, India}

\date{\today}

\begin{abstract}
We investigate the phenomenon of correlated emission lasing in a coherently driven single quantum dot coupled to a bimodal photonic crystal cavity, utilizing a master equation to describe the system dynamics. To account for exciton-phonon interactions, we incorporate a non-perturbative approach through a polaron transformed master equation. By analyzing fluctuations in the Hermitian operators associated with relative and average phase, we derive a Fokker-Planck equation to assess phase drift and diffusion coefficients, demonstrating that correlated emission suppresses quantum noise in the presence of exciton-phonon interaction at low temperature. Additionally, we calculate the single and two-photon excess emission rates (difference between emission and absorption rates) into the cavity modes and explore the generation of continuous-variable entanglement between these modes.
\end{abstract}

\maketitle

\section{Introduction}
In a correlated emission laser (CEL), coherence between the upper levels in a three-level atomic system leads to correlated spontaneous emissions into the cavity modes. This correlation suppresses quantum noise in the laser, driving it towards the vacuum noise limit (VNL). CEL has significant applications in laser gyroscopes \cite{Scully1982, Bergou1991, Mecozzi2023} and gravitational wave detectors \cite{Scully1986, Schnabel2010}, where detecting ultrasmall phase shifts in laser modes are crucial. CEL is also valuable in fields like quantum metrology, sensing and, high-resolution spectroscopy \cite{Giovannetti2006,Degen2017}, where noise from spontaneous emission often imposes limitations.

CEL has been proposed in various atomic-level configurations, such as the ‘V’ type system employed in quantum beat lasers \cite{Bergou1988QBL, Zaheer1988QBL}, Hanle lasers \cite{Bergou1988hanle, Lu1990JOSAB}, and three-level cascade systems \cite{Scully1988cascade, Lu1990cascade}. The nonlinear quantum theories for quantum beat lasers and CEL-based Hanle lasers have been developed specifically for atomic systems\cite{Lu1989,Bergou1988QBL}. CEL leads to quenching in the relative phase diffusion coefficient, which also manifests as quenching or squeezing of the quantum Hermitian operators corresponding to the relative or average phase and amplitude \cite{Lu1990JOSAB, Lu1990cascade}. Furthermore, CEL facilitates continuous variable (CV) entanglement generation between lasing modes, which provides method for generating entanglement between large number of photons\cite{Ikram2007, Ge2013}. CEL have been realized in a few remarkable experiments and a large reduction in phase diffusion noise has been observed\cite{Winters1990,Steiner1995,Abich2000}. 

With ongoing important developments in single emitter lasers, there have been some interesting proposals for single emitter CEL\cite{Kiffner2007,Dey2023}. Further, using a single superconducting three-level artificial atom, non-degenrate two-photon CEL has been realized and relative phase diffusion noise has been observed $10^{-4}$ times the Schawlow-Townes limit which is much smaller than the relative phase diffusion noise observed in CEL using ensemble of atoms\cite{Peng2015}. Thus a single emitter CEL is an ideal candidate for suppressing relative phase noise.
Realizing the CEL in semiconductor cavity QED systems is particularly exciting. As these scalable devices hold great promise for developing on-chip quantum technology due to their integrability with photonic networks. In this paper, we propose CEL using a single quantum dot (QD) embedded in a photonic bimodal cavity.

Quantum dots (QDs) exhibit discrete energy levels due to three-dimensional quantum confinement and are often referred to as 'artificial atoms'. When excited, a QD generates a quasi-particle known as an exciton, which consists of an electron-hole pair. These excitons can be either x-polarized or y-polarized. Together, the x- and y-polarized excitons and the ground state (no electron-hole pair) creates a 'V' type energy configuration. When these transitions are coupled with two orthogonally polarized cavity modes, the system resembles a Hanle laser \cite{Scully_Zubairy_1997}. At cryogenic temperatures, QD excitons primarily interact with longitudinal acoustic (LA) phonons \cite{Roy2011, Mahan1990}. These phonon interactions, which significantly influence system dynamics, are incorporated through the polaron transformed master equation. These exciton-phonon interactions lead to energy shifts, incoherent excitations that results in cavity mode feeding\cite{Hennessy2007,Kaniber2008,Senellart2009,Dalacu2010,Calic2011}, and excitation-induced dephasing phenomena\cite{Hughes2011}. We notice that it has been shown earlier that the dephasing due to relative motion of atoms in an atomic ensemble does not play significant role in noise suppression in the CEL\cite{Bergou1988hanle}.  However, it is interesting to investigate effects of exciton-phonon interactions on relative phase noise in a single QD CEL.
We consider a single quantum dot with `\textit{x}' and `\textit{y}' excitons, each driven by separate external coherent fields, and coupled to two orthogonally polarized cavity modes. Using the quantum optics toolbox\cite{SMTan1999}, we solve the master equation numerically to analyze the steady-state dynamics of the system and examine fluctuations in the relative and average phase Hermitian operators. We also calculate the phase drift and diffusion coefficients, both with and without exciton-phonon interactions (EPI), analogous to an atomic system, and derive the laser rate equations. Finally, we investigate the potential for generating continuous variable (CV) entanglement between the cavity modes.

The paper is organized as follows. In Section II, we present the results for steady-state quantum dot (QD) populations and cavity photon statistics, using the master equation for the system. We then obtain the variances of quantum Hermitian operators associated with the relative and average phase. Additionally, we derive a simplified master equation (SME) for the off-resonantly driven system, transforming it into a Fokker-Planck equation in the P-representation, and calculate the phase drift and diffusion coefficients for cases with and without exciton-phonon interactions (EPI). In Section III, we derive the laser rate equations for the system using the SME and numerically compute single and two-photon excess emissions. In Section IV, we demonstrate the potential of this system to generate continuous-variable (CV) entangled photon pairs in both cavity modes. Conclusions are provided in Section V.

\section{COHERENTLY PUMPED SINGLE QD-TWO MODE CAVITY SYSTEM}

\subsection{\label{sec:level1}Model system}

\begin{figure}
    \centering
    \includegraphics[width=\columnwidth]{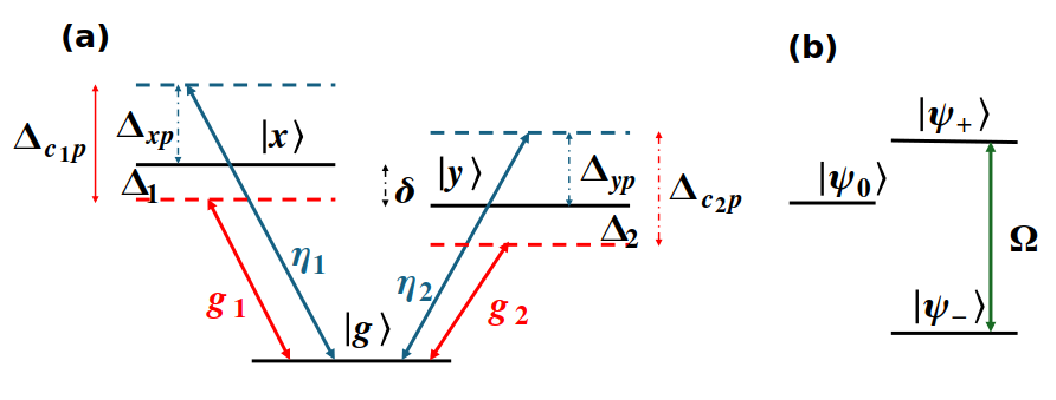}
    \caption{Schematic figure. (a) QD level structure with coherent pump and cavity mode coupling scheme. (b) Pump dressed QD states, $|\psi_+\rangle$, $|\psi_0\rangle$ and $|\psi_-\rangle$ where $\psi_\pm\rangle$ states are separated by generalized Rabi frequency, $\Omega$.}
    \label{fig:schematicFig}
\end{figure}

We consider the system having a single quantum dot coupled to bimodal cavity and $|g\rangle \leftrightarrow |x\rangle$ and $|g\rangle \leftrightarrow |y\rangle$ transitions are driven by separate coherent pumps. The schematic diagram for the system is given in Fig.\ref{fig:schematicFig}. The Hamiltonian of the system is given by,

\begin{equation}
    \begin{split}
         H = & \omega_x \sigma_1^+\sigma_1^-+\omega_y \sigma_2^+\sigma_2^-+ \omega_{c_1} a_1^\dag a_1 + \omega_{c_2} a_2^\dag a_2 
         \\&+g_1(\sigma_1^+a_1+ a_1^\dag\sigma_1^-)+g_2(\sigma_2^+a_2+ a_2^\dag\sigma_2^-)
         \\&+\eta_1(\sigma_1^+e^{-i\omega_{Lx}t}+\sigma_1^-e^{i\omega_{Lx}t})\\&+\eta_2(\sigma_2^+e^{-i\omega_{Ly}t}+\sigma_2^-e^{i\omega_{Ly}t}) + H_{ph},
    \end{split}
\end{equation}

where $a_i$ is the annihilation operator for the i-th cavity mode, $g_1$, $g_2$ are the $|g\rangle\leftrightarrow|x\rangle$, $|g\rangle\leftrightarrow|y\rangle$ transitions coupling strength to the 1st and 2nd cavity modes. QD operators are given by, $\sigma_1^+=|x\rangle\langle g|$, $\sigma_2^+=|y\rangle\langle g|$. The coherent pumping strengths are $\eta_1$, $\eta_2$ for $|g\rangle \leftrightarrow |x\rangle$, $|g\rangle \leftrightarrow |y\rangle$ transitions respectively. The exciton-phonon interaction Hamiltonian is given by, $H_{ph} = \hbar \Sigma_k\omega_k b_k^\dagger b_k + \hbar\Sigma_{i=x,y}\Sigma_k \lambda_k^i |i\rangle\langle i| (b_k^\dagger + b_k)$. Here, $b_k$ is the annihilation operator for the k-th phonon bath mode. Now, we write the above Hamiltonian in the laser frequencies rotating frame, $H'=e^{iH_0 t} H e^{-iH_0 t}$ where, $H_0=\omega_{Lx}\sigma_1^+\sigma_1^- +\omega_{Ly}\sigma_2^+\sigma_2^- +\omega_{Lx} a_1^\dag a_1 + \omega_{Ly}a_2^\dag a_2$. 

\begin{equation}
    \begin{split}
         H' = & \delta_{xp} \sigma_1^+\sigma_1^-+\delta_{yp} \sigma_2^+\sigma_2^- + \Delta_{c_1p} a_1^\dag a_1 + \Delta_{c_2p} a_2^\dag a_2
         \\&+g_1(\sigma_1^+ a_1+ a_1^\dag \sigma_1^-)+g_2(\sigma_2^+ a_2+ a_2^\dag\sigma_2^-) 
         \\&+\eta_1(\sigma_1^+ + \sigma_1^-)+\eta_2(\sigma_2^+ + \sigma_2^-) + H_{ph},
    \end{split}
\end{equation}

where, the detunings of the QD states and the cavity modes w.r.t. pump frequencies are given by $\delta_{xp}=\omega_x-\omega_{Lx}$, $\delta_{yp}=\omega_y-\omega_{Ly}=\omega_x-\delta-\omega_{L_y}$, where $\delta$ is the fine structure splitting and $\Delta_{c_1p}=\omega_{c_1}-\omega_{Lx}$, $\Delta_{c_2p}=\omega_{c_2}-\omega_{Ly}$ respectively.
To include exciton-phonon effects to all orders non-perturbatively, we make polaron transformation\cite{Mahan1990,Xu2016,Nazir2016}, $\Tilde{H}= e^S H' e^{-S}$, where $S=\Sigma_{i=1,2}\sigma_i^+\sigma_i^-\Sigma_k \frac{\lambda_k^i}{\omega_k}(b_k^\dagger-b_k)$. $\lambda_k^1$, $\lambda_k^2$ is the $|x\rangle$, $|y\rangle$ exciton and k-th phonon bath mode coupling strength respectively. The transformed Hamiltonian, $\Tilde{H}$ has form, $\Tilde{H} = H_s+H_b+H_{sb}$.

\begin{equation}
    \begin{split}
        H_s =&\hbar \Delta_{xp}\sigma_1^+\sigma_1^- + \hbar \Delta_{yp}\sigma_2^+\sigma_2^- + \\& \hbar \Delta_{c_1p}a_1^\dagger a_1+\hbar \Delta_{c_2p}a_2^\dagger a_2+\langle B\rangle X_g 
    \end{split}
\end{equation}

\begin{equation}
    H_b = \hbar \Sigma_k \omega_k b_k^\dagger b_k
\end{equation}

\begin{equation}
    H_{sb} = \zeta_g X_g + \zeta_u X_u
\end{equation}

The polaron shifts, $\Sigma_k \frac{(\lambda_k^1)^2}{\omega_k}$, $\Sigma_k \frac{(\lambda_k^2)^2}{\omega_k}$ are absorbed in the $\Delta_{xp}$, $\Delta_{yp}$. The phonon displacement operators are given by, $B_{\pm} = \exp[\pm \Sigma_{i=1,2}\Sigma_{k} \frac{\lambda_k^{i}}{\omega_k}(b_k - b_k^\dagger]$, with $\langle B_{\pm} \rangle = \langle B \rangle$. For simplification, we have considered equal coupling strengths $\lambda_k^{1} = \lambda_k^{2}$. The system operators are given by, $X_g = \hbar(g_1\sigma_1^+a + g_2\sigma_2^+a)+\hbar (\eta_1 \sigma_1^+ + \eta_2 \sigma_2^+) H.C.$, $X_u = i\hbar(g_1\sigma_1^+ a+g_2\sigma_2^+ a + \eta_1 \sigma_1^+ + \eta_2 \sigma_2^+)+H.C.$ and bath fluctuation operators are, $\zeta_g = \frac{1}{2}(B_++B_- -2\langle B \rangle)$, $\zeta_u = \frac{1}{2i}(B_+ - B_-)$. Using the polaron transformed Hamiltonian, $\Tilde{H}$ and Born-Markov approximation, we derive the master equation for the QD-cavity system \cite{Roy2011,Nazir2016}. The density matrix master equation (ME) for the system is given by,

\begin{equation}
    \begin{split}
    \dot{\rho_s} = &-\frac{i}{\hbar}[H_s,\rho_s]-L_{ph}\rho_s-\Sigma_{j=1,2}\frac{\kappa_j}{2}L[a_j]\rho_s\\&-\Sigma_{i=1,2}(\frac{\gamma_i}{2}L[\sigma_i^-]+\frac{\gamma_i'}{2}L[\sigma_i^+\sigma_i^-])\rho_s
    \end{split}
    \label{eqn:ME}
\end{equation}

where $L[\hat{O}]\rho_s = \hat{O^\dagger}\hat{O}\rho_s - 2\hat{O}\rho_s\hat{O^\dagger}+\rho_s\hat{O^\dagger}\hat{O}$ is the Lindblad superoperator. The second term in the master equation $L_{ph}\rho_s$ represents the Liouvillian capturing the effect of system-bath interaction is given by, 

\begin{equation}
    \begin{split}
        L_{ph}\rho_s = &\frac{1}{\hbar^2}\int_{0}^{\infty}d\tau \Sigma_{j=g,u}G_j(\tau)\\&\times[X_j(t),X_j(t,\tau)\rho_s(t)]+H.C.
    \end{split}
    \label{eqn:Lph}
\end{equation}

where $X_j(t,\tau)=e^{-iH_s\tau/\hbar}X_j(t)e^{iH_s\tau/\hbar}$, and polaron Green's functions, $G_j(\tau)=\langle\zeta_j(t)\zeta_j(t,\tau)\rangle_{bath}$, $G_g(\tau)=\langle B \rangle^2{\cosh(\phi(\tau)-1)}$, $G_u(\tau)=\langle B \rangle^2\sinh(\phi(\tau))$. The phonon correlation function is given by,
\begin{equation}
    \phi(\tau)=\int_{0}^{\infty}d\omega\frac{J(\omega)}{\omega^2}[\coth(\frac{\hbar\omega}{2k_BT})\cos(\omega\tau)-i\sin(\omega\tau)],
\end{equation}
\par
where $k_B$ and $T$ are the Boltzmann constant and temperature of the phonon bath, respectively. The super-ohmic spectral density function of phonon bath is given by $J(\omega)=\Sigma_k(\lambda_k^{i})^2\delta(\omega-\omega_k)=\alpha_p\omega^3\exp[-\frac{\omega^2}{2\omega_b^2}]$, takes the latter form in continuum limit\cite{Wilson2002}. In our calculations, the phonon bath parameter values considered are, electron-phonon coupling strength, $\alpha_p=2.36$ ps$^2$, and the cut-off frequency, $\omega_b=1$ meV which provide experimentally compatible values of the mean phonon displacement, $\langle B \rangle$=0.9, 0.84 and 0.73 for $T$= 5K, 10K and 20K, respectively\cite{Hughes2011}. We also include Lindblad terms corresponding to cavity damping with decay rates $\kappa_i$, spontaneous exciton decay with rate $\gamma_i$ and pure dephasing with rate $\gamma_i'$. The master equation (\ref{eqn:ME}) is numerically integrated to obtain the steady-state populations (SSP) and cavity field photon statistics. It is important to note that, in all numerical calculations, we have selected a sufficiently high photon number cutoff to ensure convergence of the results.


\subsection{\label{sec:level2} Steady state populations and cavity photon statistics}

\begin{figure}
    \centering
    \includegraphics[width=\columnwidth]{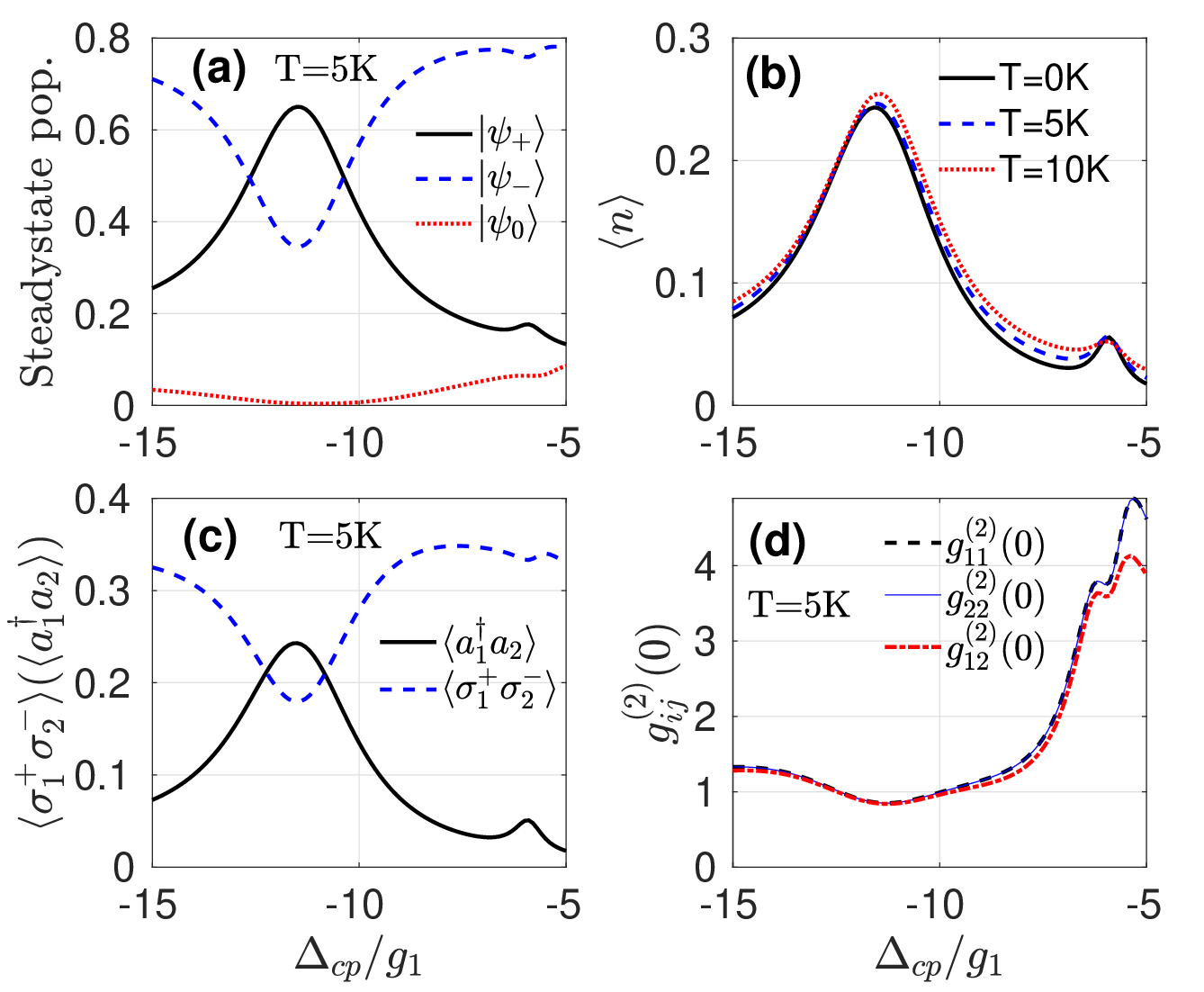}
    \caption{(a)Steady state populations of $|\psi_+\rangle$(solid black), $|\psi_-\rangle$(dashed blue) and $|\psi_0\rangle$(dotted red) for T=5K. (b) Mean cavity photon number $\langle n\rangle$ for T=0K (solid black), T=5K(dashed blue) and T=10K(dotted red). (c) Coherence between $|x\rangle$, $|y\rangle$ excitonic states, $\langle \sigma_1^+\sigma_2^-\rangle$ (dashed blue) and coherence between cavity modes, $\langle a_1^\dagger a_2\rangle$ (solid black). (d) Zero time delay second order intra and inter cavity mode correlations, $g_{11}^{(2)}(0)$ (dashed black), $g_{22}^{(2)}(0)$ (solid blue) and $g_{12}^{(2)}(0)$ (dash-dotted red) respectively. The parameters considered are QD states detuning, $\Delta_{xp}=\Delta_{yp}=\Delta=-10.0g_1$, coherent pumping rate, $\eta_1=\eta_2=\eta=2.0g_1$, cavity decay rates, $\kappa_1=\kappa_2=\kappa=0.5g_1$, spontaneous decay rate for $|x\rangle$, $|y\rangle$ states are $\gamma_1=0.01g_1$, $\gamma_2=0.01g_1$ respectively and the pure dephasing rates are $\gamma_1'=\gamma_2'=0.01g_1$.}
    \label{fig:Fig2}
\end{figure}

In Fig. \ref{fig:Fig2} we provide the results for steady state populations and cavity photon statistics. For a Hanle-type laser system, the two-photon CEL condition is given by, $|g_1|=|g_2|$, $\Delta_1=\Delta_2$ \cite{Scully1985}. Hence, we also consider QD states strongly coupled to cavity modes with equal coupling strengths $g_2=g_1=100meV$ and are off-resonant w.r.t coherent drives, $\Delta_{xp}=\Delta_{yp}=\Delta=-10.0g_1$. The exciton-phonon effects play a significant role in such off-resonantly driven systems populating the excitonic states and thereby transitions can lead to population of coupled cavity modes. We define symmetric and anti-symmetric states of the systems as $|+\rangle=\frac{|x\rangle+|y\rangle}{\sqrt{2}}$, $|-\rangle=\frac{|x\rangle-|y\rangle}{\sqrt{2}}$, assuming equal coherent pumping strengths, $\eta_1=\eta_2=\eta$. Thus, the symmetric state $|+\rangle$ is driven with strength $\sqrt{2}\eta$, leading to pump-dressed states given by $|\psi_+\rangle= \cos\alpha |+\rangle + \sin\alpha |g\rangle$, $|\psi_0\rangle=|-\rangle$, $|\psi_-\rangle=-\sin\alpha|+\rangle + \cos\alpha |g\rangle$. Here $\sin2\alpha=\sqrt{2}\eta/\Omega$, $\cos2\alpha=\Delta/\Omega$ and the generalized Rabi frequency, $\Omega=\sqrt{\Delta^2+8\times \eta^2}$. In Fig. \ref{fig:Fig2}(a) we plot the steady state populations of these pump dressed states, $|\psi_\pm\rangle$, $|\psi_0\rangle$ for the coherent pumping strengths, $\eta_2=\eta_1=\eta=2.0g_1$. We vary the cavity detunings, $\Delta_{c_1p}=\Delta_{c_2p}=\Delta_{cp}$ equally, such that $\Delta_1=\Delta_2$ satisfying the CEL condition. It is observed that for $\Delta_{cp}=-\Omega=-\sqrt{\Delta^2+8\times\eta^2}$, single-photon resonance condition, the cavity modes are tuned with the transition between the dressed states. These transitions are accompanied with the population of both the cavity modes. The mean cavity photon numbers, $\langle n_1\rangle=\langle a_1^\dagger a_1\rangle$, $\langle n_2\rangle=\langle a_2^\dagger a_2\rangle$ are given in Fig. \ref{fig:Fig2}(b) for T=5K (blue-dashed). It should be noted that both the cavity modes are equally populated. We notice that the state $|\psi_0\rangle=|-\rangle$ is decoupled from the transitions and is not populated. The presence of a small peak at $\Delta_{cp}=-\Omega/2$ corresponds to two-photon emission process. The results for the single and two-photon emission rates are provided in section \ref{sec:laserRateEq}. Further, with increase in the temperature (T), the curves are broadened and the peak at $\Delta_{cp}=-\Omega/2$ is diminished due to enhanced phonon-induced decoherence suppressing the multi-photon processes.

In Fig. \ref{fig:Fig2}(c) we show the results for the correlation between the modes, $Re(\langle a_1^\dagger a_2\rangle)$ and the coherence between upper QD states, $|x\rangle$ and $|y\rangle$ i.e., $\langle\sigma_1^+\sigma_2^-\rangle=Re(\langle y|\rho|x\rangle)$. We observe that the correlation, $\langle a_1^\dagger a_2\rangle$ rises with decrease in $\langle\sigma_1^+\sigma_2^-\rangle$ at $\Delta_{cp}=-\Omega$ implying transfer of coherence from the upper levels to the cavity modes. This correlation between the modes leads to quenching of relative or average phase fluctuations.
In Fig. \ref{fig:Fig2} (d), the results for the zero time delay photon-photon correlation functions of the modes, $g_{ij}^{(0)}=\frac{\langle a_i^\dagger a_j^\dagger a_i a_j\rangle}{\langle a_i^\dagger a_i\rangle \langle a_j^\dagger a_j\rangle}$ where $i,j={1,2}$, are given. The value of the inter-mode correlation function, $g_{12}^{(0)}$ is almost equal to that of the intra-mode correlation, $g_{ii}^{(2)}(0)$ where $i={1,2}$ at $\Delta_{cp}=-\Omega$. This shows that both the cavity modes are correlated with each other as much as with themselves.


\subsection{\label{sec:level3}Variances of the Hermitian operators}

The phase and amplitude fluctuations in the system can be evaluated by defining the Hermitian operators for the relative phase, $\phi$, the average phase, $\Phi$, the relative amplitude $r$ and the mean amplitude $R$ as given below \cite{Lu1990JOSAB},

\begin{subequations}
\begin{align}
    B_\phi &= \frac{i}{2}[a_1^\dagger e^{i\phi_1}-a_1 e^{-i\phi_1}]-\frac{i}{2}[a_2^\dagger e^{i\phi_2}-a_2 e^{-i\phi_2}]\label{subeq:Bphi}\\
    B_\Phi &= \frac{i}{4}[a_1^\dagger e^{i\phi_1}-a_1 e^{-i\phi_1}]+\frac{i}{4}[a_2^\dagger e^{i\phi_2}-a_2 e^{-i\phi_2}]\label{subeq:BPhi}\\
    B_r &= \frac{1}{2}[a_1^\dagger e^{i\phi_1} + a_1 e^{-i\phi_1}]-\frac{1}{2}[a_2^\dagger e^{i\phi_2} + a_2 e^{-i\phi_2}] \label{subeq:Br}\\
    B_R &= \frac{1}{4}[a_1^\dagger e^{i\phi_1} + a_1 e^{-i\phi_1}]+\frac{1}{4}[a_2^\dagger e^{i\phi_2} + a_2 e^{-i\phi_2}]
    \label{subeq:BR}
\end{align}
\end{subequations}

\begin{figure*}
    \centering
    \includegraphics[width=\textwidth]{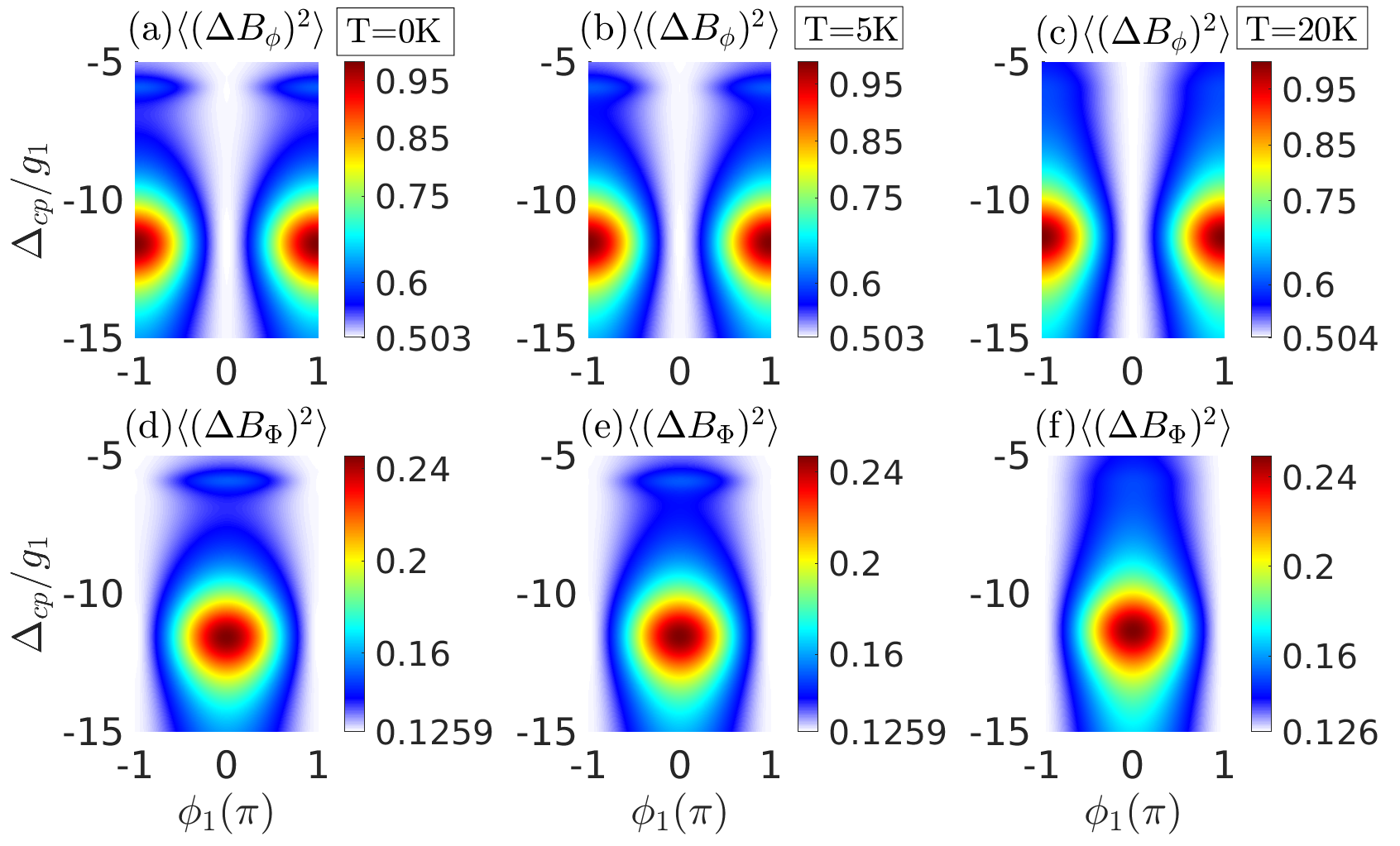}
    \caption{Variances of the Hermitian operators,  $B_\phi$ (a, b, c) and $B_\Phi$ (d, e, f) for T=0K, 5K, 20K. The other parameters are same as in Fig. \ref{fig:Fig2}(b).}
    \label{fig:Fig3}
\end{figure*}

Further, in VNL, the average value of the variances take the values, $\langle (\Delta B_{\phi,r})^2\rangle=1/2$, $\langle (\Delta B_{\Phi,R})^2\rangle=1/8$\cite{Lu1990JOSAB}. Here, we investigate the fluctuations in the relative and average phase of the cavity mode fields and the results are given in Fig. \ref{fig:Fig3}.

In Fig.\ref{fig:Fig3}, we fix the phase of second cavity mode, $\phi_2=0$ and varied the first cavity mode phase, $\phi_1$ and the cavity detuning, $\Delta_{c_1p}=\Delta_{c_2p}=\Delta_{cp}$. Fig. \ref{fig:Fig3}(b,e) shows the results for T=5K. We can see that, for $\Delta_{cp}=-\Omega$ at $\phi_1=0$, fluctuations in relative phase Hermitian operator, $B_\phi$, $\langle (\Delta B_\phi)^2\rangle\approx 0.503$ reaches VNL (0.5) implying the presence of correlated emission in the system whereas, fluctuations in the average phase, $B_\Phi$ increases above the VNL(0.125). For $\phi_1=\pm\pi$, the fluctuations in $B_\phi$ increases and in $B_\Phi$ decreases and reaches the value $\approx0.1259$. This reduction in fluctuations is manifested in the vanishing of phase diffusion coefficients shown in the following subsection. The offset in the value of variances from the VNL is attributed to phonon induced noise present in the system. We also find noise quenching at $\Delta_{cp}=-\Omega/2$, where the cavity modes are populated by two-photon emission processes. With increase in temperature, the phonon-induced incoherent scattering increases and lead to broadening and rise in the quantum fluctuations. We can compare the results for T=5K with those of T=0K $\&$ T=20K provided in Fig. \ref{fig:Fig3}(a,d) $\&$ (c,f) respectively.


\subsection{\label{sec:level 4} The phase drift and diffusion coefficients}

Further, we have obtained the drift and diffusion coefficients, $D_\phi$, $D_\Phi$, $D_{\phi\phi}$ and $D_{\Phi\Phi}$ respectively using the Fokker-Planck equation. Here, $D_\phi=0$ or $D_\Phi=0$ gives the phase locking condition and the quenching(squeezing) of the relative phase diffusion occurs when $D_{\phi\phi}=0(<0)$ and of the average phase diffusion coefficient when $D_{\Phi\Phi}=0(<0)$. 

\paragraph{Without phonons: }We derive the Fokker-Planck equation of motion for the system without including exciton-phonon interactions and calculate the average and relative phase diffusion coefficient rates following the standard procedure\cite{Scully_Zubairy_1997}. The master equation for the system ($\rho_s$) for the without exciton-phonon interaction case is given by,

\begin{equation}
 \begin{split}
  \dot{\rho_s}=& -\frac{i}{\hbar}[H_s,\rho_s]-\frac{\kappa_1}{2}L[a_1]\rho_s    -\frac{\kappa_2}{2}L[a_2]\rho_s
    \\& -\Sigma_{i=1,2}\Big[\frac{\gamma_i}{2}L[\sigma_i^-]\Big]
    \label{eqn:ME_WP}
 \end{split}
\end{equation}

Using the above equation, (\ref{eqn:ME_WP}), we obtain the master equation for the reduced density operator for cavity fields, $\rho_f$ by tracing over the quantum dot states.

\begin{equation}
    \begin{split}
    \dot{\rho_f}=& -i[\Delta_{c_1p} a_1^\dagger a_1 + \Delta_{c_2p} a_2^\dagger a_2 , \rho_f]
              -ig_1 a_1\rho_{gx} + ig_1\rho_{xg}a_1^\dagger 
              \\&-ig_2 a_2\rho_{gy}+ ig_2\rho_{yg}a_2^\dagger + ig_1\rho_{gx}a_1- ig_1 a_1^\dagger\rho_{xg} 
              \\&+ ig_2\rho_{gy}a_2- ig_2a_2^\dagger\rho_{yg}  
     \end{split}
     \label{eqn:reducedME_WP}
\end{equation}

\begin{figure}
    \centering
    \includegraphics[width=\columnwidth]{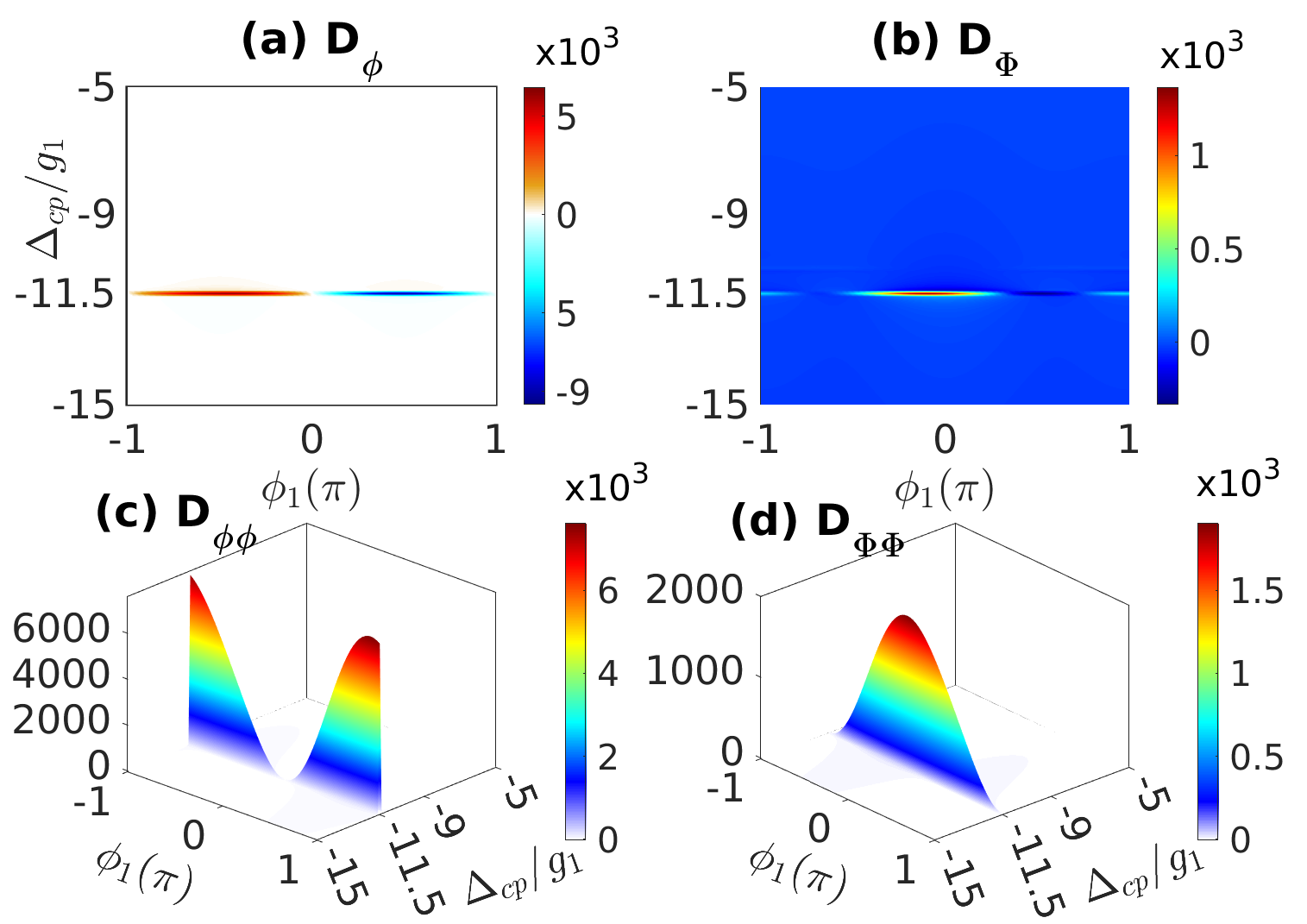}
    \caption{Without phonons: The relative and average phase drift (a,b) and diffusion (c,d) coefficients. The system parameters are same as in Fig. \ref{fig:Fig2}(b).}
    \label{fig:Fig4}
\end{figure}

The density matrix elements, $\rho_{ij}=\langle i|\rho_s|j\rangle$ are calculated and substituted back in the equation as shown in Appendix \ref{sec:AppendixB} and derived the rate equation for $\rho_f$ upto second order in the coupling strength, $g_i$. The phase diffusion coefficients are given in Fig. \ref{fig:Fig4} for the case of no exciton-phonon interactions. From Fig. \ref{fig:Fig4}(a), we can see that for the cavity detunings, $\Delta_{cp}=-\Omega$ and the phase angle, $\phi_1 = 0$, the relative drift coefficient ($D_\phi$) goes to `0'(phase locking condition) and from Fig. \ref{fig:Fig4}(c) $\&$ (d), the presence of correlated emission into the cavity modes leads to the suppression of the relative phase diffusion coefficient, $D_{\phi\phi}$ for $\phi_1 = 0$ and of the average phase diffusion coefficient, $D_{\Phi\Phi}$ for $\phi_1 = \pm\pi$ attaining VNL values.

\paragraph{With phonons:} To obtain the Fokker-Planck equation for the case that includes exciton-phonon interactions, we derive a simplified master equation(SME) for this off-resonantly driven system following the approximations, $|\Delta_{xp}|$,$|\Delta_{yp}|$, $|\Delta_{c_1p}|$, $|\Delta_{c_2p}|$ $>>g_1, g_2, \eta_1, \eta_2$. The SME gives clear view of various phonon induced processes present in this off-resonantly driven system. We write $L_{ph}\rho_s$ terms in the Lindblad form proportional to $g_i^2$, $g_ig_j$, $\eta_i^2$ and $\eta_i\eta_j$. Similar procedure has been used earlier for obtaining SME for QD-cavity QED systems\cite{Roy2011}. We also include terms not having Lindblad form such that the results of the SME converges with ME. The SME is given below,

\begin{widetext}
\begin{equation}
    \begin{split}
        \dot{\rho_s}=& -\frac{i}{\hbar}[H_{eff},\rho_s]-\frac{\kappa_1}{2}L[a_1]\rho_s    -\frac{\kappa_2}{2}L[a_2]\rho_s-\Sigma_{i=1,2}\Big[\frac{(\gamma_i+\Gamma_{ip}^+)}{2}L[\sigma_i^-] +\frac{\gamma_i'}{2}L[\sigma_i^+\sigma_i^-]+(\frac{\Gamma_i^-}{2}L[\sigma_i^+a_i]+\frac{\Gamma_i^+}{2}L[\sigma_i^-a_i^\dag]) 
        \\& + \frac{\Gamma_{ip}^-}{2}L[\sigma_i^+] \Big]\rho_s-\Big[\frac{\Gamma_{12}}{2}L[a_2^\dag\sigma_2^- ,\sigma_1^+a_1]\rho_s
        +\frac{\Gamma^p_{12}}{2}L[\sigma_1^+,\sigma_2^-]\rho_s + ( \Lambda_1^+\sigma_1^+a_1\rho_s\sigma_1^+a_1 + \Lambda_{12}^{++}\sigma_1^+a_1\rho_s \sigma_2^+a_2 
        + H.C.) \\& +\Lambda_{12}^{+-}\sigma_1^+ a_1\rho_s\sigma_2^-a_2^\dagger + (\Lambda_{1p}^+ \sigma_1^+\rho_s\sigma_1^+ + \Lambda_{12p}^{++}\sigma_1^+\rho_s\sigma_2^+ + H.C.) +\Lambda_{12p}^{+-}\sigma_1^+\rho_s\sigma_2^- + 1\leftrightarrow 2\Big]
    \label{eqn:SME}
    \end{split}
\end{equation}
\end{widetext}

Here, $L[\hat{O_1},\hat{O_2}]=\hat{O_2}\hat{O_1}\rho_s-2\hat{O_1}\rho_s\hat{O_2}+\rho_s\hat{O_1}\hat{O_2}$. The effective Hamiltonian, $H_{eff}$ signify the coherent evolution of the system. It includes the terms arising due to exciton-phonon interaction and coherent drive-phonon interaction. The phonon induced stark shifts are given by $\delta_i^{\pm}$, $\delta_{ip}^{\pm}$ and the exciton exchange-cavity photon exchange process is given by $\Omega_{12}$ and the exciton-exchange process by $\Omega_{12}^p$.

\begin{equation}
    \begin{split}
        H_{eff} = & H_S + \hbar\delta_1^+\sigma_1^+\sigma_1^- a_1a_1^\dagger + \hbar \delta_2^+\sigma_2^+\sigma_2^-a_2a_2^\dagger \\& + \hbar\delta_1^-\sigma_1^-\sigma_1^+a_1^\dagger a_1 + \hbar \delta_2^- \sigma_2^-\sigma_2^+ a_2^\dagger a_2 + \hbar \delta_{1p}^+ \sigma_1^+\sigma_1^- \\& + \hbar \delta_{2p}^+ \sigma_2^+\sigma_2^- + \hbar \delta_{1p}^- \sigma_1^-\sigma_1^+ + \hbar\delta_{2p}^- \sigma_2^-\sigma_2^+ \\& - (i\hbar \Omega_{12}\sigma_1^+\sigma_2^- a_1a_2^\dagger + i\hbar \Omega_{12}^p \sigma_1^+\sigma_2^- + H.C.)
    \end{split}
\end{equation}
 
\begin{figure}
    \centering
    \includegraphics[width=\columnwidth]{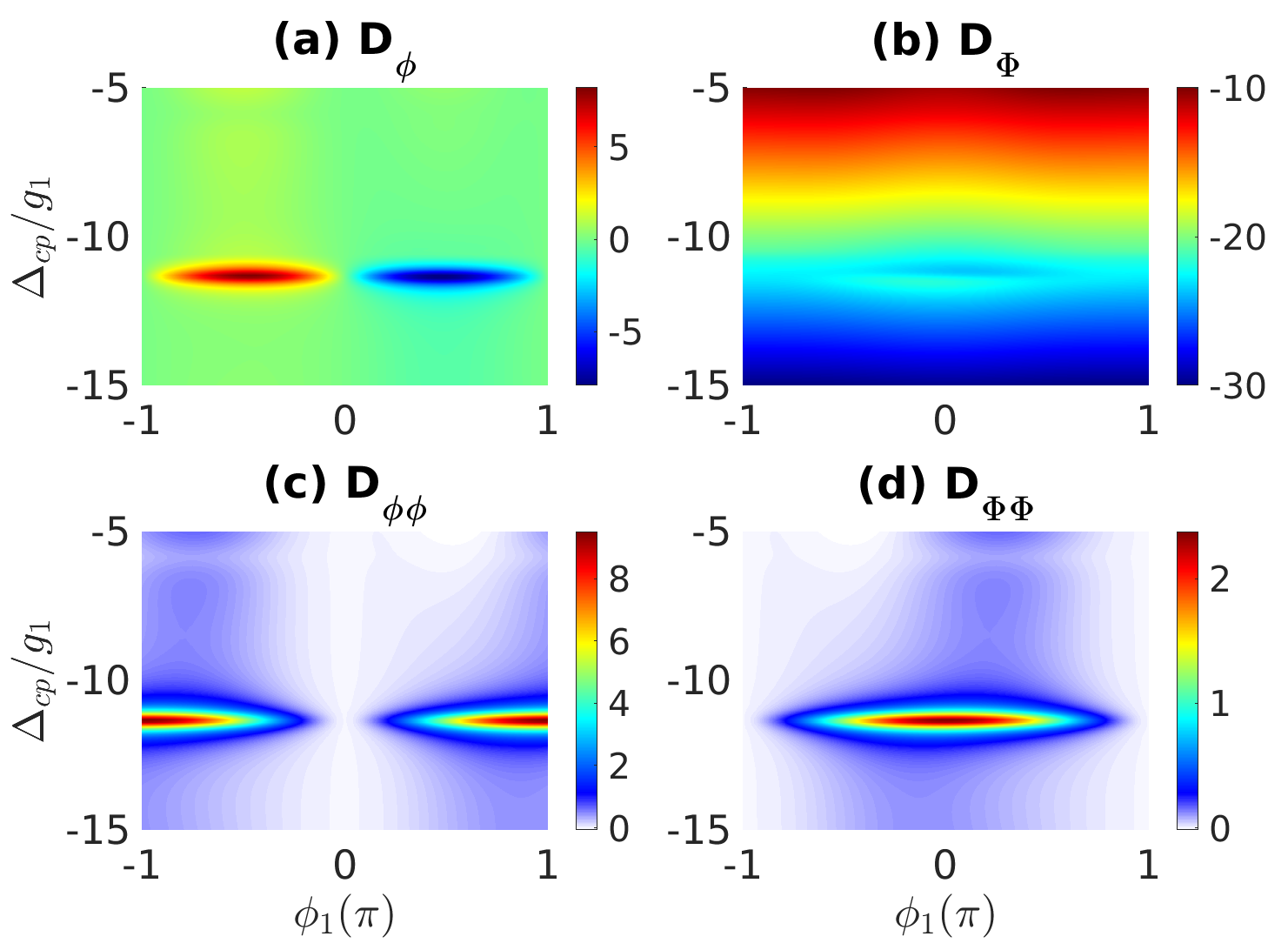}
    \caption{With phonons: The relative and average phase drift (a,b) and diffusion (c,d) coefficients for T=5K. The parameters are same as in Fig. \ref{fig:Fig2}(b).}
    \label{fig:Fig5}
\end{figure}

Additionally, the exciton-phonon interaction induces incoherent processes with scattering rates, $\Gamma_i^+$ ,($\Gamma_{i}^-)$) corresponding to QD excitation(de-excitation) by absorption(emission) of the corresponding cavity mode photon. $\Gamma_{ip}^+$ corresponding to the incoherent pumping of the QD and $\Gamma_{ip}^-$ to the enhanced radiative decay. The phonon induced incoherent exciton exchange processes rates are given by $\Gamma_{ij}$, $\Gamma_{ij}^p$. $\Lambda$'s are the coefficients for the residue terms. The form of these phonon induced scattering rates are given in Appendix \ref{sec:AppendixA}. Following a similar procedure detailed above for the without phonon case, we derive a Fokker-Planck equation for the system. Thereafter, the phase drift and diffusion coefficients are obtained. The results for T=5K are given in Fig. \ref{fig:Fig5}.

\begin{figure}
    \centering
    \includegraphics[width=\columnwidth]{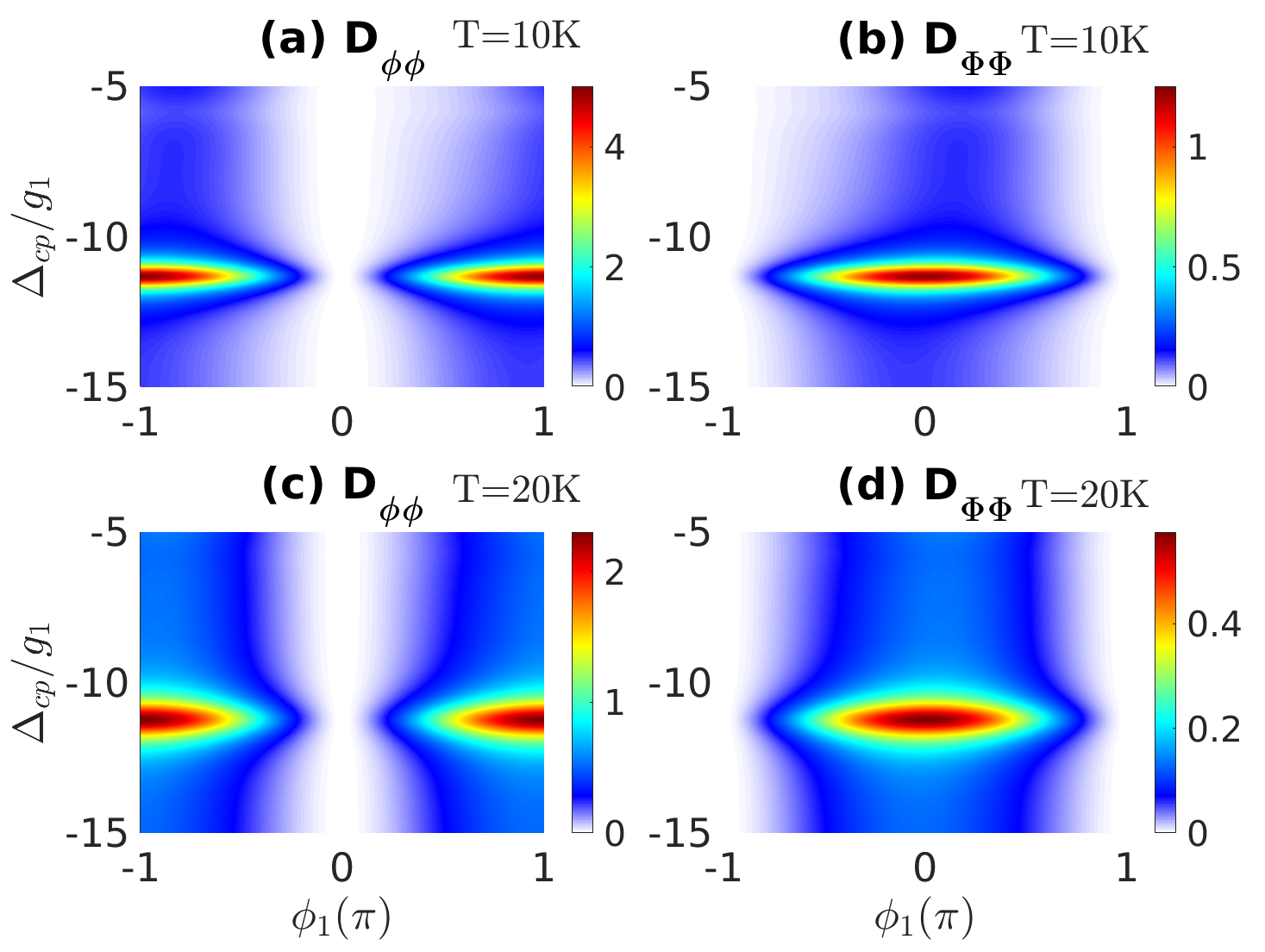}
    \caption{The relative and average phase diffusion coefficients for T=10K (a,b) and T=20K (c,d). The other parameters are same as in Fig. \ref{fig:Fig2}(b).}
    \label{fig:Fig6}
\end{figure}

We can see that correlated emission into both the cavity modes in this off-resonantly pumped single QD coupled to the bimodal photonic crystal cavity system leads to quenching in the relative or the average phase diffusion coefficients. From Fig. \ref{fig:Fig5} (a), we can see that for $\phi_1=0,\pm\pi$, at the cavity detunings, $\Delta_{cp}=-\Omega$, the relative phase drift coefficient, $D_\phi=0$ satisfies the phase locking condition. Further, the relative phase diffusion coefficient, $D_{\phi\phi}$ goes to zero at $\phi_1=0$ and the average diffusion coefficient, $D_{\Phi\Phi}=0$ at $\phi_1=\pm\pi$ showing quenching behaviour similar to the without phonon case. The correspondence with the results for variances in the Hermitian operators, $B_\phi$ $\&$ $B_\Phi$ reducing to the VNL as shown earlier (c.f. Fig. \ref{fig:Fig3}) is evident.

Fig. \ref{fig:Fig6} shows the diffusion coefficients, $D_{\phi\phi}$, $D_{\Phi\Phi}$ for T=10K in (a,b) and T=20K in (c,d). With increase in the temperature, the phonon-induced decoherence results in broadening of the diffusion coefficients implying that the cavity modes are more noisy. 


\section{\label{sec:laserRateEq}LASER RATE EQUATIONS}

In this section, we evaluate the rate of emissions into cavity mode via single and two-photon processes. To obtain the rates, we follow the standard procedure \cite{Scully1967,Sargent1974}, using the SME (\ref{eqn:SME}), we trace over the QD states and express the reduced density matrix rate equation for the cavity modes Eq.\ref{eqn:cavityRateEqn}, in terms of ``probability" flowing into and out of the state $|n,m\rangle$ where $`n'$ $\&$ $`m'$ are the number of photons in 1st $\&$ 2nd cavity modes.

\begin{equation}
    \begin{split}
        \dot{P}_{n,m} = &-\alpha_{n,m} P_{n,m}+G^{11}_{n-1,m-1}P_{n-1,m-1} 
        \\&+G^{10}_{n-1,m}P_{n-1,m}+G^{01}_{n,m-1}P_{n,m-1}
        \\&+G^{20}_{n-2,m}P_{n-2,m}+G^{02}_{n,m-2}P_{n,m-2}
        \\&+A^{11}_{n+1,m+1}P_{n+1,m+1}+A^{10}_{n+1,m}P_{n+1,m}
        \\&+A^{20}_{n+2,m}P_{n+2,m}+A^{01}_{n,m+1}P_{n,m+1}
        \\&+A^{02}_{n,m+2}P_{n,m+2} +\kappa_{1}(n+1)P_{n+1,m}- \kappa_{1}nP_{n,m} 
        \\&+\kappa_{2}(m+1)P_{n,m+1} - \kappa_{2}mP_{n,m}.
    \end{split}
    \label{eqn:cavityRateEqn}
\end{equation}

\begin{figure}
    \centering
    \includegraphics[width=\columnwidth]{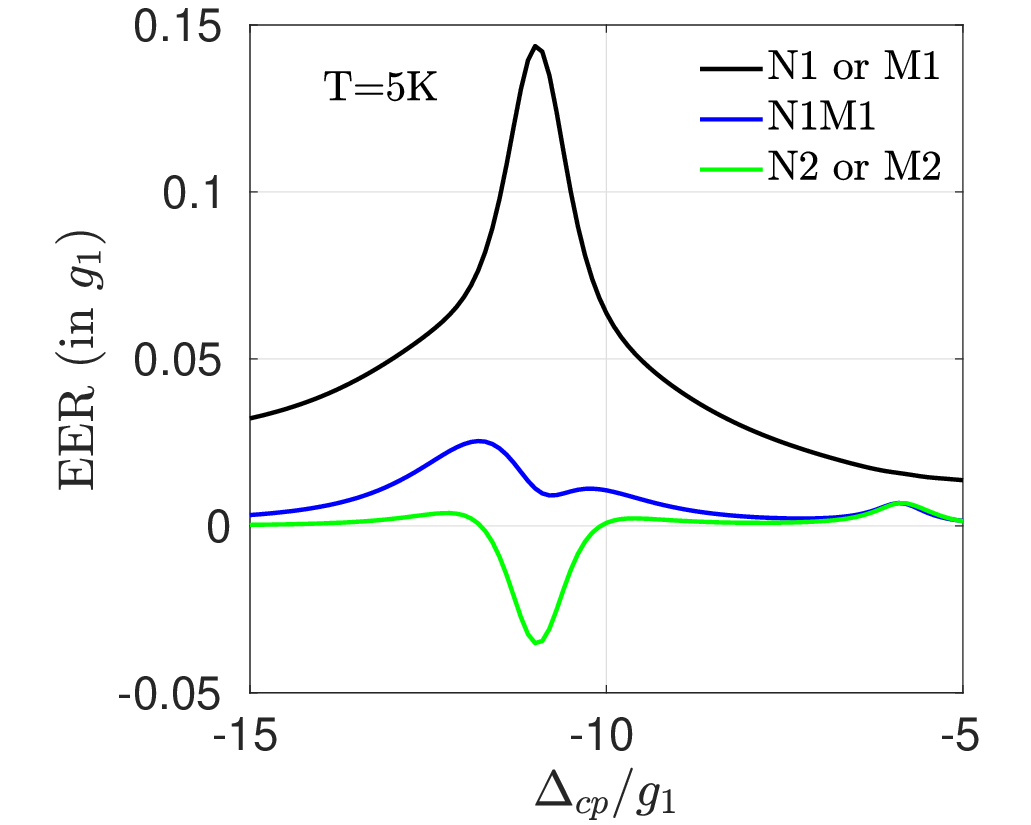}
    \caption{Single photon (first mode N1, second mode M1), two-mode two-photon (N1M1) and two-photon (first mode N2, second mode M2) excess emission rates(EER) for T=5K. All other parameters are same as in Fig.\ref{fig:Fig2}(b)}
    \label{fig:Fig7}
\end{figure}

 The probability of having $n$, $m$ photons in 1st, 2nd cavity modes is given by, $P_{nm}=\Sigma_i \langle i,n,m|\rho_s|i,n,m\rangle$, the coefficients on the right hand side are given by, $\alpha_{n,m}=\Sigma_i \alpha_{i,n,m}\langle i,n,m|\rho_s|i,n,m\rangle$, $G_{n,m}^{ab}P_{n,m}=\Sigma_i G_{i,n,m}^{ab}\langle i,n,m|\rho_s|i,n,m\rangle$ and $A_{n,m}^{ab}P_{n,m}=\Sigma_i A_{i,n,m}^{ab} \langle i,n,m|\rho_s|i,n,m\rangle$ where $i={x,y,g}$. The coefficients $\alpha_{i,n,m}$, $G_{i,n,m}^{ab}$ and $A_{i,n,m}^{ab}$ are obtained numerically. The k-photon emission(absorption) rate for the first and second modes are given by, $\Sigma_{n,m} G_{n,m}^{k0} P_{n,m}$($\Sigma_{n,m}A_{n,m}^{k0}P_{n,m}$) and $\Sigma_{n,m} G_{n,m}^{0k} P_{n,m}$($\Sigma_{n,m}A_{n,m}^{0k}P_{n,m}$) respectively and two-mode two-photon emission(absorption) rate is given by $\Sigma_{n,m} G_{n,m}^{11} P_{n,m}$($\Sigma_{n,m}A_{n,m}^{11}P_{n,m}$). The difference between emission and absorption rate is defined as `excess emission rate (EER)' and the sign of EER $>0$ or $<0$ represents net emission or absorption occurring in the cavity mode. In Fig. \ref{fig:Fig7}, we show results for the single and multi-photon excess emission rates of the system for T=5K. We can see that for $\Delta_{cp}=-\Omega=-11.5g_1$, where the fluctuations in $B_\phi$ attains VNL value and also the relative phase diffusion coefficient is quenched, there is peak in single photon emission (N1 or M1)  of both the modes and dip in other multi-photon excess emission rates. This implies photons are emitted into either of the cavity modes at equal rate and the two-photon emission and the two mode two-photon emission into the cavity modes are suppressed. Therefore, the cavity modes are populated due to single-photon emission predominantly resulting a peak in $\langle n\rangle$ c.f., Fig. \ref{fig:Fig2}(b). Also for $\Delta_{cp}=-\Omega/2$, at two-photon resonance condition, there are small peaks in both the two-mode two-photon, two-photon EER curves and a slight dip in single-photon EER curves. Further, it is noticed with increase in the temperature, rise in the phonon induced decoherence leads to the suppression in the multi-photon processes and the single photon emission dominates.


\section{CONTINUOUS VARIABLE ENTANGLEMENT BETWEEN THE CAVITY MODES}

In this section we investigate the quantum correlation between the cavity modes of the system by evaluating the Duan-Giedke-Cirac-Zoller (DGCZ) criterion\cite{DGCZ2000} for CV entanglement. We define two Einstein-Poldosky-Rosen (EPR) like variables, $u$, $v$ given by

\begin{equation}
    u=\hat{x_1}+\hat{x_2}, v=\hat{p_1}-\hat{p_2},
\end{equation}

where $\hat{x_j}$, $\hat{p_j}$ are given below in terms of cavity mode operators,
\begin{align}
    x_{j} &=\frac{1}{\sqrt{2}}(a_{j}^{\dagger}e^{-i\phi_{j}}+a_{j}e^{i\phi_{j}}),\\
    p_{j} &=\frac{i}{\sqrt{2}}(a_{j}^{\dagger}e^{-i\phi_{j}}-a_{j}e^{i\phi_{j}}),j=\{1,2\}.
\end{align}

According to DGCZ criterion, the sum of the variances of $u$, $v$ i.e., $\Delta u^2 + \Delta v^2 = (\langle u^2-\langle u\rangle^2\rangle)^2+ (\langle v^2-\langle v\rangle^2\rangle)^2 < 2$ is the sufficient condition for the presence of CV entanglement between the modes. 

\begin{figure}
    \centering
    \includegraphics[width=\columnwidth]{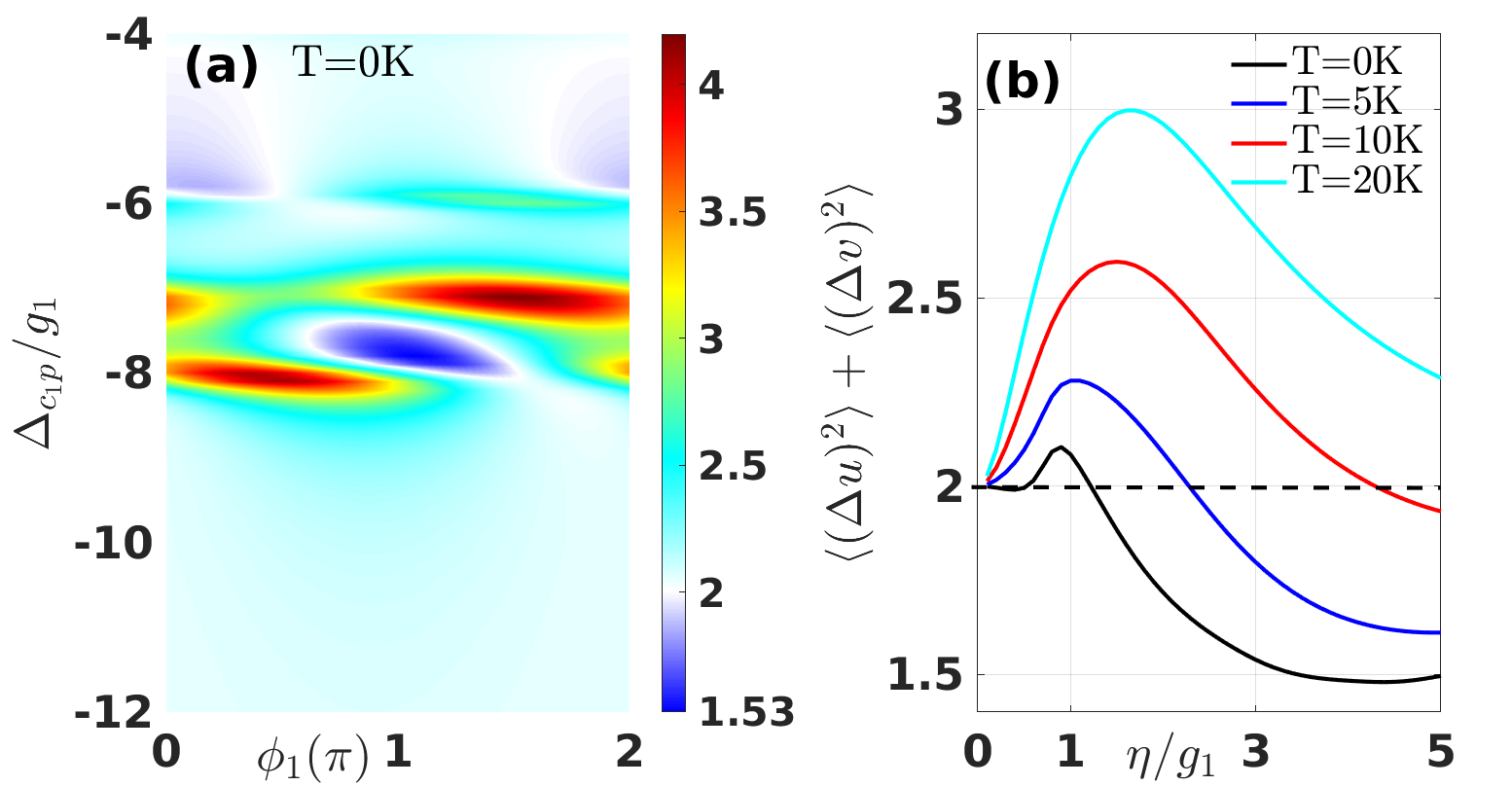}
    \caption{Entanglement criterion, $(\Delta u)^2 + (\Delta v)^2$. (a) Varying $\Delta_{c_1p}$ and $\phi_1$ for T=0K. (b) Varying coherent pumping strength $\eta$ for $\Delta_{c_1p}=-\Omega$, $\phi_1=\pi$. Detunings, $\Delta_{xp}=\Delta_{yp}=\Delta=5.0g_1$, $\kappa=0.1g_1$, $\phi_2=0$ and the other parameters are same as in Fig. \ref{fig:Fig2}(b).}
    \label{fig:Fig8}
\end{figure}

The entanglement between the cavity modes can be witnessed for two-photon condition, $\Delta_{c_1p}+\Delta_{c_2p}=0$. We consider that the QD-states are positively detuned w.r.t the coherent pump, $\Delta_{xp}=\Delta_{yp}=\Delta=5.0g_1$ such that the phonon induced decoherence is small compared to $\Delta<0$ case, preserving the entanglement between the cavity modes. It is observed that the total variance, $(\Delta u)^2 + (\Delta v)^2$ attains lowest value when the cavity detunings are $\pm\Omega$ such that $\Delta_{c_1p}+\Delta_{c_2p}=0$. In Fig. \ref{fig:Fig8}(a), we fix the detuning of 2nd cavity mode, $\Delta_{c_2p}=\Omega$ and vary the detuning of the 1st cavity mode, $\Delta_{c_1p}$ and $\phi_1$. We can see that for $\phi_1=\pi$ and $\Delta_{c_1p}=-\Omega$, the total variance with its minimum value, $(\Delta u)^2 + (\Delta v)^2=1.525$, the DGCZ criterion for the CV entanglement is satisfied implying quantum correlation between the cavity modes. It should be noted that the EPR variables have the form similar to that of $B_R$, $B_\phi$ c.f. Eq. \ref{subeq:BR}, Eq. \ref{subeq:Bphi} respectively. Therefore we can say, squeezing in either of the operators, $B_R$, $B_\phi$ or both as in cascaded system \cite{Lu1990cascade} leads to the entanglement between the cavity modes. Therefore, the inter-mode two-photon correlation is greater than the intra-mode correlation. In Fig. \ref{fig:Fig8}(b) we show the change in total variance with increase in temperature and coherent pumping rate by fixing the cavity detunings, $\Delta_{c_1p}=-\Omega$, $\Delta_{c_2p}=\Omega$ and $\phi_1=\pi$. With increase in pumping rate, the cavity modes get populated and the entanglement between them is generated. This is attributed to the domination of the two-mode two-photon excess emission into the cavity mode over other processes. Further increase in pumping rate leads to increased single photon EER  and thereby diminishing entanglement. It is clear that at higher temperature phonon induced incoherent processes dominate and the cavity modes are disentangled. We also noticed that the entanglement between the modes is reduced with increase in cavity decay rates and is lost completely.  Hence, the CV entanglement between the cavity modes in this coherently-pumped strongly coupled system is maintained at low temperatures.


\section{CONCLUSIONS}

We have derived a master equation for a coherently pumped single quantum dot (QD) coupled to a bimodal cavity system and investigated its steady-state dynamics, as well as the fluctuations in the relative and average phase Hermitian operators. A Fokker-Planck equation was derived for this system, revealing that the relative and average phase diffusion coefficients vanish due to correlated emission into the cavity modes. This feature suggests that such a semiconductor system could serve as a viable platform for on-chip cavity-enhanced light (CEL). Furthermore, we calculated the single- and two-photon emission rates into the cavity modes using a simplified master equation (SME) and examined the generation of continuous variable (CV) entanglement between the cavity modes.


\appendix
\section{\label{sec:AppendixA}PHONON INDUCED SCATTERING RATES}

The phonon induced scattering rates are given below,

\begin{equation}
    \delta_{i}^{\pm} = g_i^2 Im\Big[\int_0^\infty d\tau G_+e^{\pm i\Delta_i\tau}\Big]
\end{equation}

\begin{equation}
    \delta_{ip}^{\pm} = \eta_i^2 Im\Big[\int_0^\infty d\tau G_+e^{\pm i\Delta_{c_ip}\tau}\Big]
\end{equation}

\begin{equation}
    \Omega_{ij} = \frac{g_ig_j}{2}\int_0^\infty d\tau(G_+e^{ i\Delta_j\tau}-G_+^*e^{-i\Delta_i\tau})
\end{equation}

\begin{equation}
    \Omega_{ij}^p = \frac{\eta_i\eta_j}{2}\int_0^\infty d\tau(G_+e^{ i\Delta_{jp}\tau}-G_+^*e^{-i\Delta_{ip}\tau})
\end{equation}

\begin{equation}
    \Gamma_i^{\pm}=g_i^2\int_0^\infty d\tau (G_+ e^{\pm i\Delta_i\tau}+G_+^* e^{\mp i\Delta_i\tau)}
\end{equation}

\begin{equation}
    \Gamma_{ip}^{\pm}=\eta_i^2\int_0^\infty d\tau (G_+ e^{\pm i\Delta_{c_ip}\tau}+G_+^* e^{\mp i\Delta_{c_ip}\tau})
\end{equation}

\begin{equation}
    \Gamma_{ij} = g_ig_j\int_0^\infty d\tau(G_+e^{ i\Delta_i\tau}+G_+^*e^{-i\Delta_j\tau})
\end{equation}

\begin{equation}
    \Gamma_{ij}^p = \eta_i\eta_j\int_0^\infty d\tau(G_+e^{ i\Delta_{ip}\tau}+G_+^*e^{-i\Delta_{jp}\tau})
\end{equation}

\begin{equation}
    \Lambda_i^\pm = g_i^2\int_0^\infty d\tau (G_- + G_-^*)e^{\mp i\Delta_i\tau})
\end{equation}

\begin{equation}
    \Lambda_{ij}^{\pm\pm} = g_ig_j\int_0^\infty d\tau (G_-e^{\mp i\Delta_i\tau} + G_-^*e^{\mp i\Delta_j\tau})
\end{equation}

\begin{equation}
    \Lambda_{ij}^{+-} = g_ig_j\int_0^\infty d\tau (G_- e^{-i\Delta_i\tau}+G_-^* e^{i\Delta_j\tau})
\end{equation}

\begin{equation}
    \Lambda_{ip}^{\pm} = \eta_i^2\int_0^\infty d\tau (G_- + G_-^*)e^{\mp i\Delta_{ip}\tau}
\end{equation}

\begin{equation}
    \Lambda_{ijp}^{\pm\pm} = \eta_i\eta_j\int_0^\infty (G_- e^{\mp i\Delta_{ip}\tau} + G_-^* e^{\mp i\Delta_{jp}\tau})
\end{equation}

\begin{equation}
    \Lambda_{ijp}^{+-} = \eta_i\eta_j\int_0^\infty (G_-e^{-i\Delta_{ip}\tau} + G_-^* e^{i\Delta_{jp}\tau})
\end{equation}

\section{\label{sec:AppendixB}DERIVATION OF DRIFT AND DIFFUSION COEFFICIENTS}

For the without phonons case, we obtain the density matrix elements, $\rho_{ij} = \langle i|\rho_s|j\rangle$ upto zeroth  in coupling strength, $g_i$ by solving the matrix rate equation given by, $\dot{R^{(0)}}=-MR^{(0)}+X$,  Here, $R=\begin{bmatrix}\rho_{xg} & \rho_{yg} & \rho_{gx} & \rho_{gy} & \rho_{xx} & \rho_{yy} & \rho_{xy} & \rho_{yx}\end{bmatrix}^T$ , $X=\begin{bmatrix}i\eta_1 & i\eta_1 & -i\eta_2 & -i\eta_2 & 0 & 0 & 0 & 0\end{bmatrix}^T$ 

\begin{widetext}
$M=\begin{bmatrix}
-i\Delta_{xp}-\frac{\gamma_1}{2} & 0 & 0 & 0 & 2i\eta_1 & i\eta_1 & i\eta_2 & 0\\
0 & -i\Delta_{yp}-\frac{\gamma_2}{2} & 0 & 0 & i\eta_2 & 2i\eta_2 & 0 & i\eta_1\\
0 & 0 & i\Delta_{xp}-\frac{\gamma_1}{2} & 0 & -2i\eta_1 & -i\eta_1 & 0 & -i\eta_2\\
0 & 0 & 0 & i\Delta_{yp}-\frac{\gamma_1}{2} & -i\eta_2 & -2i\eta_2 & -i\eta_1 & 0\\
i\eta_1 & 0 & -i\eta_1 & 0 & -\gamma_1 & 0 & 0 & 0\\
0 & i\eta_2 & 0 & -i\eta_2 & 0 & -\gamma_2 & 0 & 0\\
i\eta_2 & 0 & 0 & -i\eta_1 & 0 & 0 & -i(\Delta_{xp}-\Delta_{yp})-\frac{\gamma_1+\gamma_2}{2} & 0\\
0 & i\eta_1 & -i\eta_2 & 0 & 0 & 0 & 0 & i(\Delta_{xp}-\Delta_{yp})-\frac{\gamma_1+\gamma_2}{2}
\end{bmatrix}$

and 1st order in $g_i$ are obtained by integrating the rate equation, $\dot{R}=-MR-ig_1A_1e^{i\Delta_{c_1p}t}-ig_2A_2e^{i\Delta_{c_2p}t}-ig_1A_3e^{-i\Delta_{c_1p}t}-ig_2A_4e^{-i\Delta_{c_2p}t}$.

\begin{align}
    A1 = \begin{bmatrix} ig_1\rho_{xx}^{(0)}a_1-ig_1a_1\rho_{gg}^{(0)} \\ ig_1\rho_{yx}^{(0)}a_1 \\ 0 \\ 0 \\ -ig_1a_1\rho_{gx}^{(0)} \\ 0 \\ -ig_1a_1\rho_{gy}^{(0)}\\ 0 \end{bmatrix}; 
    A2 = \begin{bmatrix} ig_2\rho_{xy}^{(0)}a_2 \\ ig_2\rho_{yy}^{(0)}a_2 - ig_2a_2\rho_{gg}^{(0)} \\ 0 \\ 0 \\ 0 \\ -ig_2a_2\rho_{gy}^{(0)} \\ 0 \\ -ig_2a_2\rho_{gx}^{(0)} \end{bmatrix} ; \\
    A3 = \begin{bmatrix} 0 \\ 0 \\ -ig_1a_1^\dagger \rho_{xx}^{(0)} + ig_1\rho_{gg}^{(0)}a_1^\dagger \\ -ig_1a_1^\dagger \rho_{xy}^{(0)} \\ ig_1\rho_{xg}^{(0)}a_1^\dagger \\ 0 \\ 0 \\ ig_1\rho_{yg}^{(0)}a_1^\dagger\end{bmatrix};
    A4 = \begin{bmatrix} 0 \\ 0 \\ -ig_2a_2^\dagger\rho_{yx} \\ -ig_2a_2^\dagger\rho_{yy}^{(0)} + ig_2\rho_{gg}^{(0)}a_2^\dagger \\ 0 \\ ig_2\rho_{yg}^{(0)}a_2^\dagger \\ ig_2\rho_{xg}^{(0)}a_2^\dagger \\ 0 \end{bmatrix}
\end{align}

\begin{equation}
    \begin{split}
        R^{(1)}(t) = & (M-i\Delta_{c_1p})^{-1}A1 + (M-i\Delta_{c_2p})^{-1}A2 +(M+i\Delta_{c_1p})^{-1}A3 + (M+i\Delta_{c_2p})^{-1}A4
    \end{split}
\end{equation}

After substituting the matrix elements, $\rho_{ij}^{(1)}$ into the Eq. (7), the reduced density matrix for the cavity field takes the following form,

\begin{equation}
    \begin{split}
         \dot{\rho_f} = & -i\Delta_{c_1p}a_1^\dagger a_1\rho_f + i\Delta_{c_1p}\rho_f a_1^\dagger a_1 -i\Delta_{c_2p}a_2^\dagger a_2\rho_f +i\Delta_{c_2p}\rho_f a_2^\dagger a_2 \\&+ \Big[\alpha_{11}\rho_f a_1a_1^\dagger - \alpha_{11}a_1^\dagger \rho_f a_1 + \alpha_{12}\rho_f a_2a_1^\dagger - \alpha_{12}a_1^\dagger \rho_f a_2 \\&+ \alpha_{13}\rho_f a_1^\dagger a_1^\dagger - \alpha_{13}a_1^\dagger \rho_f a_1^\dagger + \alpha_{14}\rho_f a_2^\dagger a_1^\dagger - \alpha_{14}a_1^\dagger \rho_f a_2^\dagger \\& + \alpha_{15}a_1 \rho_f a_1^\dagger - \alpha_{15} a_1^\dagger a_1\rho_f + \alpha_{16}a_2 \rho_f a_1^\dagger - \alpha_{16} a_1^\dagger a_2\rho_f \\& + \alpha_{17}a_1^\dagger \rho_f a_1^\dagger - \alpha_{17} a_1^\dagger a_1^\dagger\rho_f + \alpha_{18}a_2^\dagger \rho_f a_1^\dagger - \alpha_{18} a_1^\dagger a_2^\dagger\rho_f
         \\&+ \nu_{21}\rho_f a_1a_2^\dagger - \nu_{21}a_2^\dagger \rho_f a_1 + \nu_{22}\rho_f a_2a_2^\dagger - \nu_{22}a_2^\dagger \rho_f a_2 \\&+ \nu_{23}\rho_f a_1^\dagger a_2^\dagger - \nu_{23}a_2^\dagger \rho_f a_1^\dagger + \nu_{24}\rho_f a_2^\dagger a_2^\dagger - \nu_{24}a_2^\dagger \rho_f a_2^\dagger \\& + \nu_{25}a_1 \rho_f a_2^\dagger - \nu_{25} a_2^\dagger a_1\rho_f + \nu_{26}a_2 \rho_f a_2^\dagger - \nu_{26} a_2^\dagger a_2\rho_f \\& + \nu_{27}a_1^\dagger \rho_f a_2^\dagger - \nu_{27} a_2^\dagger a_1^\dagger\rho_f + \nu_{28}a_2^\dagger \rho_f a_2^\dagger - \nu_{28} a_2^\dagger a_2^\dagger\rho_f + H.C.\Big]
    \end{split}
\end{equation}

the coefficients, $\alpha_{ij},\nu_{ij}$ are obtained numerically after the substitution of $\rho_{ij}^{(1)}$ . The Fokker-Planck equation in the P-representation is given below after making the substitutions,  $a_i\rho_f \rightarrow q_iP$; $\rho_f a_i^\dagger \rightarrow q_i^*P$; $a_i^\dagger\rho_f \rightarrow (q_i^*-\frac{\partial}{\partial q_i})P$; $\rho_f a_i\rightarrow (q_i-\frac{\partial}{\partial q_i^*})P a_i^\dagger a_i\rho_f \rightarrow (q_i^*-\frac{\partial}{\partial q_i})$.  

\begin{equation}
    \begin{split}
        \frac{\partial P}{\partial t} = & \Big[ (\Delta_{c_1p}+\alpha_{11}+\alpha_{15})\frac{\partial}{\partial q_1}q_1 + (\Delta_{c_2p}+\nu_{22}+\nu_{26})\frac{\partial}{\partial q_2}q_2 +  H.C.\Big]P
        \\& + \Big[ (\alpha_{13}+\alpha_{17})\frac{\partial}{\partial q_1}q_1^* + (\nu_{24}+\nu_{28})\frac{\partial}{\partial q_2}q_2^* + H.C. \Big] P
        \\& + \Big[ (\alpha_{12}+\alpha_{16})\frac{\partial}{\partial q_1}q_2 + (\nu_{21}+\nu_{25})\frac{\partial}{\partial q_2}q_1 + H.C. \Big] P
        \\& + \Big[ (\alpha_{14}+\alpha_{18})\frac{\partial}{\partial q_1}q_2^* + (\nu_{23}+\nu_{27})\frac{\partial}{\partial q_2}q_1^* + H.C. \Big] P
        \\& - \Big[ \alpha_{17}\frac{\partial^2}{\partial q_1^2} + \nu_{28}\frac{\partial^2}{\partial q_2^2} + \alpha_{18}\frac{\partial^2}{\partial q_1\partial q_2} + \nu_{27}\frac{\partial^2}{\partial q_2\partial q_1} + H.C. \Big]
        \\& - \Big[ \alpha_{11}\frac{\partial^2}{\partial q_1\partial q_1^*} + \nu_{22}\frac{\partial^2}{\partial q_2\partial q_2^*} + \alpha_{12}\frac{\partial^2}{\partial q_1\partial q_2^*} + \nu_{21}\frac{\partial^2}{\partial q_2\partial q_1^*} + H.C. \Big]P
    \end{split}
    \label{eqn:FokkerPlanckEq}
\end{equation}

Further, defining $q_i$ in polar coordinates, $q_i = r_i e^{i\phi_i}$, the average and relative phase are given by $\phi=\phi_1-\phi_2$ and $\Phi=\frac{\phi_1+\phi_2}{2}$ respectively. We also have,

\begin{equation}
    \frac{\partial}{\partial q_l} = \frac{e^{-i\phi_l}}{2}(\frac{\partial}{\partial r_l} + \frac{1}{ir_l}\frac{\partial}{\partial \phi_l})
\end{equation}

\begin{eqnarray}
    \frac{\partial}{\partial \phi_1} = \frac{1}{2}\frac{\partial}{\partial \Phi}+\frac{\partial}{\partial \phi}\\
    \frac{\partial}{\partial \phi_2} = \frac{1}{2}\frac{\partial}{\partial \Phi}-\frac{\partial}{\partial \phi}
\end{eqnarray}

Considering negligible variations in the mean photon number, we get

\begin{equation}
    \frac{\partial}{\partial q_l} = \frac{e^{-i\phi_l}}{2}(\frac{1}{ir_l}\frac{\partial}{\partial \phi_l})
\end{equation}

After rewriting the Eq. \ref{eqn:FokkerPlanckEq} in terms of average and relative phases ($\Phi$ and $\phi$), 

\begin{equation}
    \frac{\partial P}{\partial t} = \frac{\partial}{\partial \phi} D_\phi P + \frac{\partial}{\partial \Phi} D_\Phi P + \frac{\partial^2}{\partial \theta^2}D_{\phi\phi} P + \frac{\partial^2}{\partial \Phi^2}D_{\Phi\Phi}P + \frac{\partial^2}{\partial \phi \partial \Phi} D_{\phi\Phi}P
\end{equation}

The drift and the diffusion coefficients are given by, 

\begin{equation}
    \begin{split}
        D_\phi = & \Delta_{c_1p}-\Delta_{c_2p}\\&-\frac{i}{2}\Big[ (\alpha_{11}-\nu_{22}) + (\frac{\alpha_{12}r_2}{r_1}e^{-i\phi}-\frac{\nu_{21}r_1}{r_2}e^{i\phi})-(\frac{\alpha_{12}}{r_1r_2}e^{-i\phi}-\frac{\nu_{21}}{r_1r_2}e^{i\phi})+(\alpha_{13}e^{-i(2\Phi+\phi)}-\nu_{24}e^{-i(2\Phi-\phi)})
        \\&+(\alpha_{14}\frac{r_2}{r_1}e^{-i2\Phi}-\nu_{23}\frac{r_1}{r_2}e^{-i2\Phi})+(\alpha_{15}-\nu_{26})+(\alpha_{16}\frac{r_2}{r_1}e^{-i\phi}-\nu_{25}\frac{r_1}{r_2}e^{i\phi})+(\alpha_{17}e^{-i(2\Phi+\phi)}-\nu_{28}e^{-i(2\Phi-\phi)})\\&-(\frac{3\alpha_{17}}{2r_1^2}e^{-i(2\Phi+\phi)}-\frac{3\nu_{28}}{2r_2^2}e^{-i(2\Phi-\phi)})+(\alpha_{18}\frac{r_2}{r_1}e^{-i2\Phi}-\nu_{27}\frac{r_1}{r_2}e^{-i2\Phi}) \Big] + H.C.
    \end{split}
\end{equation}

\begin{equation}
    \begin{split}
        D_\Phi = & \frac{\Delta_{c_1p}}{2}+\frac{\Delta_{c_2p}}{2}
        \\& -\frac{i}{4}\Big[ (\alpha_{11}+\nu_{22})+(\alpha_{12}\frac{r_2}{r_1}e^{-i\phi} + \nu_{21}\frac{r_1}{r_2}e^{i\phi}) + (\alpha_{13}e^{-i(2\Phi+\phi)}+\nu_{24}e^{-i(2\Phi-\phi)}) + (\alpha_{14}\frac{r_2}{r_1}e^{-i2\Phi}+\nu_{23}\frac{r_1}{r_2}e^{-i2\Phi})
        \\& + (\alpha_{15}+\nu_{26})+(\alpha_{16}\frac{r_2}{r_1}e^{-i\phi}+\nu_{25}\frac{r_1}{r_2}e^{i\phi})+(\alpha_{17}e^{-i(2\Phi+\phi)}+\nu_{28}e^{-i(2\Phi-\phi)})-(\frac{3\alpha_{17}}{2r_1^2}e^{-i(2\Phi+\phi)}+\frac{3\nu_{28}}{2r_2^2}e^{-i(2\Phi-\phi)})
        \\& + (\alpha_{18}\frac{r_2}{r_1}e^{-i2\Phi}+\nu_{27}\frac{r_1}{r_2}e^{-i2\Phi})-(\frac{\alpha_{18}}{r_1r_2}e^{-i2\Phi}+\frac{\nu_{27}}{r_1r_2}e^{-i2\Phi}) \Big] + H.C.
    \end{split}
\end{equation}

\begin{equation}
    D_{\phi\phi} = \frac{1}{4}\Big[ -(\frac{\alpha_{11}}{r_1^2}+\frac{\nu_{22}}{r_2^2})+(\frac{\alpha_{12}}{r_1r_2}e^{-i\phi}+\frac{\nu_{21}}{r_1r_2}e^{i\phi})+(\frac{\alpha_{17}}{r_1^2}e^{-i(2\Phi+\phi)}+\frac{\nu_{28}}{r_2^2}e^{-i(2\Phi-\phi)})-(\frac{\alpha_{18}}{r_1r_2}e^{-i2\Phi}+\frac{\nu_{27}}{r_1r_2}e^{-i2\Phi})\Big]+H.C.
\end{equation}

\begin{equation}
    D_{\Phi\Phi} = \frac{1}{16}\Big[ -(\frac{\alpha_{11}}{r_1^2}+\frac{\nu_{22}}{r_2^2})-(\frac{\alpha_{12}}{r_1r_2}e^{-i\phi}+\frac{\nu_{21}}{r_1r_2}e^{i\phi})+(\frac{\alpha_{17}}{r_1^2}e^{-i(2\Phi+\phi)}+\frac{\nu_{28}}{r_2^2}e^{-i(2\Phi-\phi)})+(\frac{\alpha_{18}}{r_1r_2}e^{-i2\Phi}+\frac{\nu_{27}}{r_1r_2}e^{-i2\Phi})\Big]+H.C.
\end{equation}

\end{widetext}

\bibliography{CEL}

\end{document}